\documentclass[aps,nofootinbib]{revtex4}
\usepackage{epsfig,graphics,amssymb,amsmath,subeqnarray,color}
\def\d{{\rm d}}

\begin{document}

\title{The viscous friction of a mesh-like super-hydrophobic surface}
\title{The  friction of a mesh-like super-hydrophobic surface}
\author{Anthony M. J. Davis\footnote{Email: amdavis@ucsd.edu}}
\author{Eric Lauga\footnote{Email: elauga@ucsd.edu}}
\affiliation{Department of Mechanical and Aerospace Engineering, 
University of California San Diego, 
9500 Gilman Drive, La Jolla CA 92093-0411, USA.}
\date{\today}
\begin{abstract}
When a liquid droplet is located above a super-hydrophobic surface, it only barely touches the solid portion of the surface, and therefore slides very easily on it.  More generally, super-hydrophobic surfaces have been shown to lead to significant reduction of viscous friction in the laminar regime, so it is of interest to quantify their effective slipping properties as a function of their geometric characteristics. Most previous studies have considered {flows bounded} by {arrays of either long grooves,  or isolated solid pillars on an otherwise flat solid substrate}, and for which therefore the surrounding air constitutes the continuous phase. Here we consider instead the  case where the super-hydrophobic surface is made of isolated holes in an otherwise continuous no-slip surface, and specifically focus on the mesh-like geometry recently  achieved experimentally. We present  an analytical method to calculate the friction of  such a surface in the case where the mesh is thin. The results for the effective slip length of the surface are computed, compared to simple estimates, and a practical fit is proposed displaying a logarithmic dependance on the {area} fraction of the solid surface.

\end{abstract}
\maketitle

\section{Introduction}

\begin{figure}[t]
\begin{center}
\includegraphics[width=0.25\textwidth]{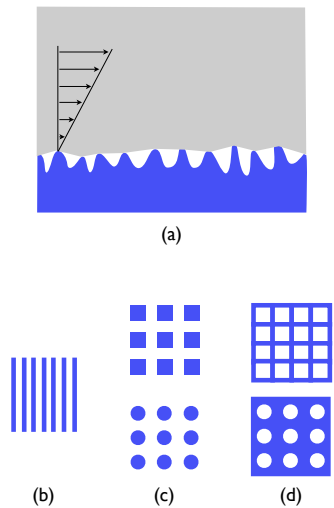}
\end{center}
\caption{(color online) Typical  topology of super-hydrophobic surfaces:
(a) Side view of a random  super-hydrophobic surface, illustrating the {contact of the liquid (gray) at the edge of the surface roughness (blue)};
(b) Top view of a {one-dimensional} surface;
(c) Top view of regular {two-dimensional} surfaces where the air (white) is the continuous phase;
(d) Top view of regular {two-dimensional} surfaces where the solid (blue) is the continuous phase. The case illustrated on  Fig.~\ref{general}d (top) is the focus of the current paper.}
\label{general}
\end{figure}

Among the fascinating flow phenomena occurring on small scales \cite{stone_review,squiresquake}, super-hydrophobicity offers  a unique bridge between microscopic features 
and macroscopic  behavior    \cite{feng02,quere05,sun05,quere08}. Super-hydrophobic surfaces are non-wetting surfaces which possess sufficiently-large geometrical roughness that a liquid droplet {deposited on the surface
would not fill the grooves of the surface roughness, but instead remain in a   a fakir-like state where the droplet only  touches the surface at the edge of the  roughness }(Fig.~\ref{general}a). As a result, super-hydrophobic surfaces possess very high effective contact angles, and  display remarkable macro-scale wetting properties \cite{feng02,quere05,sun05,quere08}.

One particularly interesting characteristic of super-hydrophobic surfaces is their low viscous friction. Since fluid in contact with the surface only barely touches it, but is instead mostly in contact with the surrounding air, small droplets can roll very easily, a phenomenon known as the lotus-leaf effect. In general, super-hydrophobic surfaces are expected to provide opportunities for significant drag reduction in the  laminar regime, as has been confirmed by  experiments cited below. The effective viscous friction of a solid surface is usually quantified by a so-called slip length, denoted here by $\lambda$, which is the distance below the solid surface where the no-slip boundary would be satisfied if the flow field was linearly extrapolated, and the no-slip boundary condition corresponds to  $\lambda =0$   \cite{neto05,laugareview,bocquet07}.

At low Reynolds number, the only characteristics affecting the friction of super-hydrophobic surfaces arise from  their geometry, specifically (a) the distribution of liquid/solid and liquid/air contact at the edge of the roughness elements of the surface, and (b) the shape  of the liquid/air free surface. For {one-dimensional} surfaces, {\it i.e.} surfaces which have one homogeneous direction (Fig.~\ref{general}b), significant drag reduction can be obtained when the homogeneous direction is parallel or perpendicular to the direction of flow, as demonstrated experimentally \cite{ou04,ou05,gogte05,choi06,truesdell06,maynes07,lee08} and theoretically \cite{philip72,philip72_2,lauga03,cottin-bizonne04,davies06,maynes07,ybert07,ng09}.

For {two-dimensional} surfaces,  an additional free parameter is the topology of the air/solid partition on the planar surface, and two general types can be distinguished. In the first type, the air is the continuous phase, and the liquid is in contact with the solid only at isolated, unconnected, locations (Fig.~\ref{general}c). This is the most commonly-studied type of super-hydrophobic surface, and arises for  surface roughness in the shape of bumps or posts on an otherwise flat solid surface  \cite{ou04,ou05,gogte05,choi06_PRL,joseph06,ybert07,lee08,ng09_2}. The second type of surface, less studied, is one where the solid is the continuous phase, and the liquid/air contact occurs on isolated domains (Fig.~\ref{general}d). These surfaces can be obtained by making holes  in an otherwise flat material \cite{ng09_2}, and have been used to study the influence of the geometry of the  liquid/air free interface on the viscous friction of the surface \cite{steinberger07,hyvaluoma08,legendre08_2,davis09}. 

\begin{figure}[t]
\begin{center}
\includegraphics[width=0.45\textwidth]{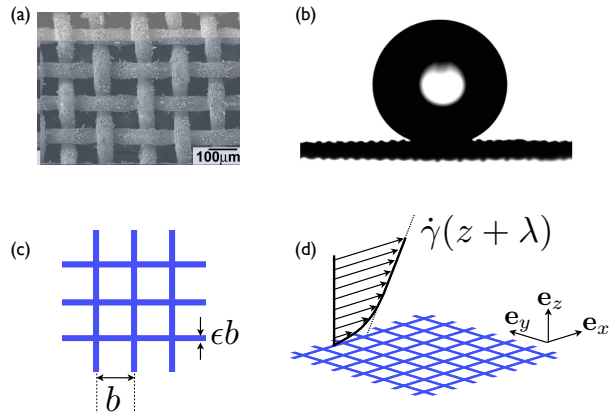}
\end{center}
\caption{(color online) Mesh-like super-hydrophobic surface. 
(a): Coated copper mesh at the micron scale \cite{wang07}; 
(b): The mesh from Figs.~\ref{grid}a  is super-hydrophobic (water droplet, volume about 4 $\mu$l)   \cite{wang07};
(c): For our calculation we consider a planar square  mesh of periodicity $b$ and width $\epsilon b$;
(d): A shear flow is set up in one of the principal directions of the mesh, with shear rate $\dot{\gamma}$; the far-field flow is given by ${\bf v}=\dot\gamma (z + \lambda) {\bf e}_x$, where $\lambda$ is the effective slip length of the mesh.
Figs.~\ref{grid}a and \ref{grid}b reprinted from S. Wang, Y. Song and L. Jiang  (2007) 
``Microscale and nanoscale hierarchical structured mesh films with 
super-hydrophobic and superoleophilic 
properties induced by long-chain fatty 
acids'',  {\it Nanotechnology}, {\bf  18},  015103, courtesy of Shutao Wang and with permission from IOP Publishing. 
}
\label{grid}
\end{figure}

A super-hydrophobic surface with a continuous solid phase can also be obtained by using an intertwined solid mesh. Recently this  method has been  exploited experimentally,  using coated steel \cite{feng04,sun05}, and coated copper  \cite{wang07}, as reproduced in Fig.~\ref{grid}a and b. The resulting surface is also the relevant geometrical configuration for flow past textile and fabric material.  In this paper, we present  an analytical method to calculate the friction of  such a mesh-like (or grid-like)  surface in the case where the mesh is thin in the plane (mesh aspect ratio $\epsilon \ll 1$).   {The analysis, based on a distribution of flow singularities,  leads to an infinite system of linear equations for the approximate Fourier coefficients of the flow around the mesh, and gives results with relative error of order $\epsilon \ln (1/\epsilon)$}. 
 The results for the effective slip length of the surface are  computed, compared to simple estimates,  and used to propose a practical fit displaying the expected  logarithmic dependance on the solid {area} fraction \cite{philip72,philip72_2,lauga03,ybert07}.

\section{Shear flow past a rectangular grid}

\subsection{Calculation background}
\label{history}

Our paper follows classical work on the broadside motion of rigid bodies at low Reynolds number
 in the asymptotic limit for narrow cross-sections \cite{happel,kimbook}. Here, we build 
on the method of Leppington and Levine  who studied axisymmetric potential problems involving an  annular disk by using a distribution of singularities, and obtained an efficient method in the limit where the radii difference of the disk approaches zero \cite{leppington72}.  Roger and Hussey studied the flat annular ring problem  both experimentally and theoretically using a beads-on-a-shell model to represent a distribution of point forces \cite{roger82}.  Subsequently,  Stewartson  showed that viscous fluid exhibits a marked reluctance to flow through a thin torus and the flux function has an essential singularity in the limit in which the hollow boundary approaches a circle and disappears \cite{stewartson83}.  This `blockage' feature  was further demonstrated by Davis and James in considering 
flow through a  array of narrow annular disks placed in planes normal to  the flow in order to model a fibrous
medium  \cite{davis96}.  Their analysis was made tractable by retaining only the inverse square root term in
the force density function.  A similar asymptotic estimate was used by
Davis in modeling the broadside oscillations of a thin grid
of the type discussed below  \cite{davis93}. 
Since edgewise motions induce the
same edge singularity in the force density, {we follow in this paper a similar analytical treatment}.

\subsection{Model problem}

We consider a thin stationary planar square mesh in the $(x,y)$ plane subject to a shear flow, with shear rate $\dot\gamma$,  along the $x$ direction, which is one of the principal directions of the mesh, as illustrated in Figs.~\ref{grid}c and d. The mesh has periodicity $b$ and a small width $\epsilon b$. We denote by ${\bf e}_x$ and ${\bf e}_y$ the unit vectors parallel to the mesh axes, and ${\bf e}_z$ is the third Cartesian vector. We further assume that the shape of the interface between the liquid and the air located underneath the mesh is planar, with no protrusion of the mesh inside the fluid. Since the mesh is a super-hydrophobic surface and presents to the flow a combination of  no-slip domains (the mesh) and perfectly-slipping domains (the free surfaces), the effective flow past the surface can be described as a slipping flow, and is given   in the far field  by ${\bf v}=\dot\gamma (z + \lambda) {\bf e}_x$, where $\lambda$ is the effective surface slip length ($\lambda > 0$). The goal of the calculation below is to present an accurate analytic calculation for $\lambda$  in the limit where $\epsilon\ll 1$.

In order to perform the calculation, we move in the frame translating at the steady velocity $U{\bf e}_x$, with $  U=\dot\gamma \lambda$. In that case, the problem is equivalent to  the uniform translation of an thin mesh in its own plane at velocity $-U{\bf e}_x$, and the goal is to derive the value of the shear rate $\dot\gamma$ in the far-field, $|z| \gg b$. The slip length will then be given by $\lambda = U/\dot\gamma$.
Edge effects are ignored by assuming an infinite mesh as then periodic point force singularities can be used to  describe the fluid flow field. The  rectangular grid naturally introduces Fourier series with respect to Cartesian coordinates, and the analysis establishes a set of integral equations with the same logarithmic dependence on the geometrical parameters of the grid as in the flows described {in the references discussed in \S\ref{history}.}

\begin{figure}[t]
\begin{center}
\includegraphics[width=0.3\textwidth]{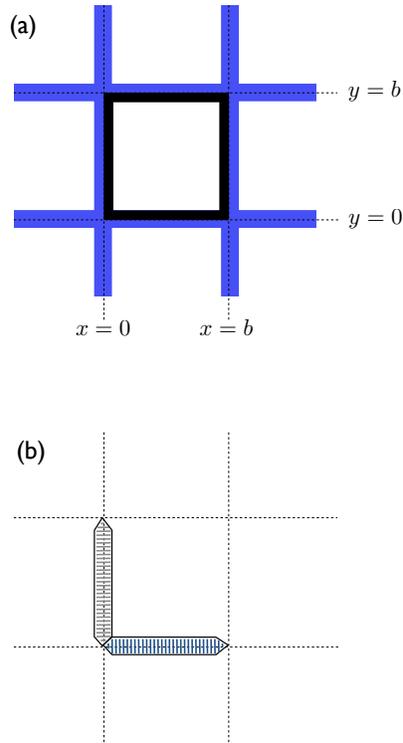}
\end{center}
\caption{(color online)  A typical grid element (see text for details).}
\label{zoom}
\end{figure}

The Reynolds number of the viscous incompressible flow is assumed to be
sufficiently small for the velocity field ${\bf v}$ to satisfy the
creeping flow (Stokes) equations \cite{happel,kimbook}
\begin{equation}\label{stokes}
\mu\nabla^2{\bf v}=\nabla p, \qquad \nabla\cdot{\bf v}=0,  
\end{equation}
where $\mu$ is the coefficient of viscosity and $p$ is the dynamic 
pressure. The prescribed grid velocity is given by 
\begin{equation}\label{Gvel}
{\bf v}=-U{\bf e}_x \hbox{ at }z=0,  \qquad (x,y)\in G.
\end{equation}
In a typical grid element $0\leq x,y\leq b$, $G$ is complementary to the square hole bounded by the lines $x$ or
$y=\frac{1}{2}\epsilon b$ or $(1-\frac{1}{2}\epsilon) b$ (shown in black in Fig.~\ref{zoom}a). 

\subsection{Solution using a superposition of singularities}
{Following the calculation for broadside oscillations of the grid in Ref.~\cite{davis93},}
the fluid motion can be represented as due to a distribution of
tangentially directed Stokeslets over the flat mesh $G$ and the density
functions must be both periodic in two dimensions and symmetric with
respect to the sides of each square. The field due to a two-dimensional square
array, period $b$, of point forces of strength $4\pi\mu Ub$ directed
parallel to the mesh's motion, is governed by
\begin{equation}\label{div}
\nabla\cdot{\bf v}_A=0
\end{equation}
and
\begin{eqnarray}\label{forcearray}
\mu\nabla^2{\bf v}_A-\nabla p_A&=&4\pi\mu Ub {\bf e}_x\delta(z)
\sum_{n_1=-\infty}^\infty\sum_{n_2=-\infty}^\infty\delta(x-n_1b)
\delta(y-n_2b)\nonumber\\
&=&\frac{4\pi\mu U}{b} {\bf e}_x \delta(z)\sum_{m_1=-\infty}^\infty \sum_{m_2=-\infty}^\infty\exp\left[\frac{2\pi i}{b}(m_1x+m_2y)\right].   
\end{eqnarray}
With ${\bf m}=m_1{\bf e}_x+m_2{\bf e}_y$, the ${\bf m}={\bf 0}$
term in Eqs.~\eqref{div} and (\ref{forcearray}) yields
\begin{equation}
\mu\nabla^2{\bf v}_0-\nabla p_0=\frac{4\pi\mu U}{b}{\bf e}_x
\delta(z), \qquad  \nabla\cdot{\bf v}_0=0,
\end{equation}
whose solution, 
\begin{equation}
{\bf v}_0=\frac{2\pi U|z|}{b}{\bf e}_x, \qquad p_0=0, 
\end{equation}
exhibits the anticipated shear at infinity. Note that the suppression
of the immaterial arbitrary multiple of $U{\bf e}_x$ in ${\bf v}_0$
ensures uniqueness below. The solution of
\begin{equation}
\mu\nabla^2{\bf v}-\nabla p=\frac{2\mu U}{b}{\bf e}_x
\int_{-\infty}^\infty e^{ikz}dk\sum_{{\bf m}}{}'\exp\left[\frac
{2\pi i}{b}{\bf m}\cdot{\bf r}\right], \qquad  \nabla\cdot{\bf v}=0, 
\end{equation}
with a prime denoting that the ${\bf m}={\bf 0}$ term is omitted, is
readily found by Fourier transform techniques (see Ref.~\cite{hasimoto59}). 
Thus the flow governed by Eqs.~\eqref{div} and (\ref{forcearray}) is compactly expressed as
\begin{equation}\label{arrayflow}
{\bf v}_A=U\left[\left(\frac{2\pi|z|}{b}-S_1\right){\bf e}_x+
\nabla\frac{\partial S_2}{\partial x}\right], \qquad 
p_A=\mu U\frac{\partial S_1}{\partial x},
\end{equation}
where
\begin{equation}\label{ess1}
S_1=\sum_{{\bf m}}{}'\frac{1}{|{\bf m}|}\exp\left[\frac
{2\pi}{b}(i{\bf m}\cdot{\bf r}-|{\bf m}||z|)\right]=\nabla^2 S_2,
\end{equation}
with
\begin{eqnarray}\label{ess2}
S_2&=&-\frac{2}{b}\int_{-\infty}^\infty e^{ikz}dk\sum_{{\bf m}}{}'
\exp\left[\frac{2\pi i}{b}{\bf m}\cdot{\bf r}\right]
\left[\left(\frac{2\pi i}{b}{\bf m}\right)^2+k^2\right]^{-2}\nonumber\\
&=&-\frac{b^2}{8\pi^2}\sum_{{\bf m}}{}'\frac{1}{|{\bf m}|^2}\left(
\frac{1}{|{\bf m}|}+\frac{2\pi}{b}|z|\right)\exp\left[\frac
{2\pi}{b}(i{\bf m}\cdot{\bf r}-|{\bf m}||z|)\right]\cdot
\end{eqnarray}
Only the velocities at the mesh and at infinity are needed for the
subsequent analysis.  The solution Eq.~(\ref{arrayflow}) shows that
\begin{equation}\label{veeAinf}
{\bf v}_A\sim\frac{2\pi U|z|}{b}{\bf e}_x { ={\bf v}_0} \hbox{ as }|z|\to\infty,
\end{equation}
{since all terms in Eq.~\eqref{ess1} exhibit exponential decay,} 
and
\begin{equation}\label{veeAgrid}
[{\bf v}_A]_{z=0}=U{\bf e}_x\left(-S_1+\frac{\partial^2S_2}
{\partial x^2}\right)_{z=0}=-U{\bf e}_x\sum_{{\bf m}}{}'
\frac{1}{|{\bf m}|}
\exp\left[\frac{2\pi i}{b}{\bf m}\cdot{\bf r}\right]C_{{\bf m}},
\end{equation} 
where, after substitution of Eqs.~(\ref{ess1}) and (\ref{ess2}), 
we have 
\begin{equation}\label{ceem}
C_{{\bf m}}=1-\frac{m_1^2}{2|{\bf m}|^2}\cdot
\end{equation}
In particular, we see that
\begin{subeqnarray}\label{ceeminf}
C_{(m_1,0)}&=&\frac{1}{2}, \qquad C_{{\bf m}}\to\frac{1}{2} \hbox{ as } m_1\to\infty, \\
C_{(0,m_2)}&=&1, \qquad C_{{\bf m}}\to 1 \hbox{ as }m_2\to\infty. 
\end{subeqnarray}

The shaded region in Fig.~\ref{zoom}b suffices for the distribution of periodic
point forces and the enforcement of the prescribed mesh velocity 
(Eq.~\ref{Gvel}). The hexagon in Fig.~\ref{zoom}b  lying along the $x$-axis is given by
$|y|\leq\frac{1}{2}\epsilon b,|y|\leq x\leq b-|y|$, or
$-1\leq w,s\leq 1$ in terms of new variables $(w,s)$ defined by
\begin{equation}\label{newvars}
w=\frac{\frac{1}{2}b-x}{\frac{1}{2}b-y}, \qquad y=\frac{1}{2}\epsilon 
bs, \qquad x=\frac{1}{2}b[1-(1-\epsilon s)w].
\end{equation}
Here $w$ may be identified as the tangent of an angle subtended at the
center of the square. The hexagon in Fig.~\ref{zoom}b  lying along the $y$-axis is given 
similarly by interchanging $x$ and $y$ in Eq.~(\ref{newvars}). The flow
generated by the translating mesh {can be written as}
\begin{eqnarray} \label{vee}
{\bf v}&=&\frac{1}{4}\int_{-1}^1\int_{-1}^1\left\{f_x(w,s){\bf v}_A
\left[x-\frac{1}{2}b(1-w+\epsilon sw),y-\frac{1}{2}\epsilon bs,z\right]
\right.\nonumber\\
&& \left. +f_y(w,s){\bf v}_A\left[x-\frac{1}{2}\epsilon bs,
y-\frac{1}{2}b(1-w+\epsilon sw),z\right]\right\}\,\d w\d s,
\end{eqnarray}
{where $f_x$ and $f_y$ are  dimensionless force densities.} The prescribed mesh velocity is then obtained by enforcing Eq.~(\ref{Gvel}) at
each point of the two hexagons. Thus Eq.~(\ref{vee}) gives
\begin{subeqnarray}\label{veegrid}
-U{\bf e}_x &=& \frac{1}{4}\int_{-1}^1\int_{-1}^1\left\{f_x(w,s)
{\bf v}_A\left[\frac{1}{2}b(w-W+\epsilon SW-\epsilon sw),
\frac{1}{2}\epsilon b(S-s),0\right]\right. \nonumber\\
&& \left. + f_y(w,s){\bf v}_A\left[\frac{1}{2}b(1-W+\epsilon SW-
\epsilon s),\frac{1}{2}b(\epsilon S-1+w-\epsilon sw),0\right]\right\}
\,\d w\d s, \quad \\
-U{\bf e}_x &=& \frac{1}{4}\int_{-1}^1\int_{-1}^1\left\{f_x(w,s)
{\bf v}_A\left[\frac{1}{2}b(\epsilon S-1+w-\epsilon sw),
\frac{1}{2}b(1-W+\epsilon SW-\epsilon s),0\right]\right. \nonumber \\
&& \left. + f_y(w,s){\bf v}_A\left[\frac{1}{2}\epsilon b(S-s),
\frac{1}{2}b(w-W+\epsilon SW-\epsilon sw),0\right]\right\}\,\d w\d s 
\end{subeqnarray}
for  all $ -1\leq W,S\leq 1$.  We further note that the summation in Eq.~(\ref{veeAgrid}) takes the simpler form
\begin{eqnarray}
\sum_{{\bf m}}{}'\frac{1}{|{\bf m}|}
\exp\left[\frac{2\pi i}{b}{\bf m}\cdot{\bf r}\right]C_{{\bf m}}=
\sum_{m_1=1}^\infty \frac{1}{m_1}\cos\frac{2\pi}{b}m_1x + 2
\sum_{m_2=1}^\infty \frac{1}{m_2}\cos\frac{2\pi}{b}m_2y \nonumber\\
+4\sum_{m_1=1}^\infty\sum_{m_2=1}^\infty \frac{1}{\sqrt{m_1^2+m_2^2}}
\left[1-\frac{m_1^2}{2(m_1^2+m_2^2)}\right]
\cos\frac{2\pi}{b}m_1x \cos\frac{2\pi}{b}m_2y, 
\end{eqnarray}
from which we note that large contributions to Eqs.~(\ref{veegrid}) arise
from summations of type
\begin{equation}
\sum_{m=1}^\infty \frac{1}{m}\cos\pi m\epsilon(S-s){ = \ln\left[\frac{1}{2}\csc \frac{\pi \epsilon}{2}|S-s|\right]}
\sim 
-\ln(\pi\epsilon|S-s|).
\end{equation}
Substitution of Eq.~(\ref{veeAgrid}) into Eqs.~(\ref{veegrid}) and use 
of Eqs.~(\ref{ceeminf}) now yields, when terms that tend to zero as 
$\epsilon\to 0$ are neglected,
\begin{subeqnarray}\label{inteq2D}
1 &=& \frac{1}{4}\int_{-1}^1\int_{-1}^1\left\{f_x(w,s)\left[
\sum_{m_1=1}^\infty\frac{1}{m_1}\cos\pi m_1(w-W)-2\ln(\pi\epsilon|S-s|)
\right.\right. \nonumber\\
&&\left. +4\sum_{m_1=1}^\infty\cos\pi m_1(w-W)\left\langle\sum_{m_2=1}
^\infty\left(\frac{1}{|{\bf m}|}C_{{\bf m}}-\frac{1}{m_2}\right)-
\ln(\pi\epsilon|S-s|)\right\rangle\right] \nonumber \\
&&+f_y(w,s)\left[\sum_{m_1=1}^\infty \frac{(-1)^{m_1}}{m_1}\cos\pi m_1W
+2\sum_{m_2=1}^\infty \frac{(-1)^{m_2}}{m_2}\cos\pi m_2w\right. \nonumber \\
&& \left.\left.+4\sum_{m_1=1}^\infty\sum_{m_2=1}^\infty 
\frac{(-1)^{m_1+m_2}}{|{\bf m}|}C_{{\bf m}}\cos\pi m_1W\cos\pi m_2w
\right]\right\}\,\d w\d s,\\
1 &=& \frac{1}{4}\int_{-1}^1\int_{-1}^1\left\{f_x(w,s)\left[
\sum_{m_1=1}^\infty \frac{(-1)^{m_1}}{m_1}\cos\pi m_1w+
2\sum_{m_2=1}^\infty\frac{(-1)^{m_2}}{m_2}\cos\pi m_2W\right.\right. \nonumber \\
&&\left.+4\sum_{m_1=1}^\infty\sum_{m_2=1}^\infty \frac{(-1)^{m_1+m_2}}
{|{\bf m}|}C_{{\bf m}}\cos\pi m_1w\cos\pi m_2W\right] \nonumber\\
&&+f_y(w,s)\left[-\ln(\pi\epsilon|S-s|)+2\sum_{m_2=1}^\infty
\frac{1}{m_2}\cos\pi m_2(w-W)\right. \nonumber \\
&&\left.\left. +4\sum_{m_2=1}^\infty\cos\pi m_2(w-W)\left\langle
\sum_{m_1=1}^\infty\left(\frac{1}{|{\bf m}|}C_{{\bf m}}-\frac{1}{2m_1}
\right)-\frac{1}{2}\ln(\pi\epsilon|S-s|)\right\rangle\right]\right\}
\,\d w\d s\quad\quad\,\,
\end{subeqnarray}
for all 
$-1\leq W,S\leq 1$.  All the $w$-integrals in these integral equations can be expressed in 
terms of Fourier coefficients defined as 
\begin{equation}\label{Fcoeffts}
[f_{xn}(s),f_{yn}(s)]=\frac{1}{2}\int_{-1}^1 [f_x(w,s),f_y(w,s)]
\cos n\pi w \d w \qquad (n\geq 0),
\end{equation}
because the symmetric forcing ensures that the corresponding sine
coefficients are all zero, {and where we use brackets to define simultaneously two sets of Fourier coefficients.}
We then identify Eqs.~(\ref{inteq2D}) as a pair
of Fourier cosine series in $W$, for each $S$ in $[-1,1]$, and, by
considering coefficients of $\cos\pi m_1W$ and $\cos\pi m_2W$ in the
respective series, we obtain 
\begin{subeqnarray}\label{inteqs}
1=-\int_{-1}^1 f_{x0}(s)\ln(\pi\epsilon|S-s|)ds+\sum_{m_2=1}^\infty
\frac{(-1)^{m_2}}{m_2}\int_{-1}^1 f_{ym_2}(s)\d s, \\
0=\int_{-1}^1 f_{xm_1}(s)\left[\frac{1}{2m_1}+2\sum_{m_2=1}^\infty
\left(\frac{1}{|{\bf m}|}C_{{\bf m}}-\frac{1}{m_2}\right)-2
\ln(\pi\epsilon|S-s|)\right]\d s\nonumber\\
+\frac{(-1)^{m_1}}{2m_1}\int_{-1}^1f_{y0}(s)ds+2\sum_{m_2=1}^\infty 
\frac{(-1)^{m_1+m_2}}{|{\bf m}|}C_{{\bf m}}\int_{-1}^1f_{ym_2}(s) \d s, 
\qquad (m_1\geq 1), \\
2=-\int_{-1}^1 f_{y0}(s)\ln(\pi\epsilon|S-s|)ds+\sum_{m_1=1}^\infty
\frac{(-1)^{m_1}}{m_1}\int_{-1}^1 f_{xm_1}(s)\d s, \\
0=\int_{-1}^1 f_{ym_2}(s)\left[\frac{1}{m_2}+\sum_{m_1=1}^\infty
\left(\frac{2}{|{\bf m}|}C_{{\bf m}}-\frac{1}{m_1}\right)-
\ln(\pi\epsilon|S-s|)\right]\d s \nonumber\\
+\frac{(-1)^{m_2}}{m_2}\int_{-1}^1f_{x0}(s)ds+2\sum_{m_1=1}^\infty 
\frac{(-1)^{m_1+m_2}}{|{\bf m}|}C_{{\bf m}}\int_{-1}^1f_{xm_1}(s)\d s, 
\qquad (m_2\geq 1),
\end{subeqnarray} 
for all $-1\leq S\leq 1$. These two related sets of equations have the form
\begin{equation}\label{ABeqn1n}
\int_{-1}^1 [f_{xn}(s),f_{yn}(s)]\ln\left(\frac{1}{|S-s|}\right)\d s
=\hbox{ constant }=[A_n,B_n]\ln 2  \qquad (-1\leq S\leq 1), 
\end{equation}
whose solution is
\begin{equation}
[f_{xn}(s),f_{yn}(s)]=\frac{[A_n,B_n]}{\pi(1-s^2)^{1/2}}  
\qquad (-1\leq s\leq 1),
\end{equation}
with the particular property
\begin{equation}\label{ABeqn2n}
\int_{-1}^1 [f_{xn}(s),f_{yn}(s)]\d s=[A_n,B_n].
\end{equation}
Thus, as demonstrated more rigorously by Leppington and Levine \cite{leppington72}
and exploited frequently elsewhere, the logarithmic kernel obtained as
an asymptotic estimate yields the inverse square root function as the
associated asymptotic solution. The substitution of Eqs.~(\ref{ABeqn1n}) and
(\ref{ABeqn2n}) into Eqs.~(\ref{inteqs}) finally yields an infinite system of linear equations 
\begin{subeqnarray}\label{ABeqn3n}
1&=&A_0\ln\left(\frac{2}{\pi\epsilon}\right)+\sum_{m_2=1}^\infty
\frac{(-1)^{m_2}}{m_2}B_{m_2}, \slabel{a}\\
0&=&A_{m_1}\left[\ln\left(\frac{2}{\pi\epsilon}\right)+\frac{1}{4m_1}+
\sum_{m_2=1}^\infty\left(\frac{1}{|{\bf m}|}C_{{\bf m}}-\frac{1}{m_2}
\right)\right] \nonumber\\
&& +\frac{(-1)^{m_1}}{4m_1}B_0+\sum_{m_2=1}^\infty\frac{(-1)^{m_1+m_2}}
{|{\bf m}|}C_{{\bf m}}B_{m_2}, 
\quad (m_1\geq 1),\\
2 & = & B_0\ln\left(\frac{2}{\pi\epsilon}\right)+\sum_{m_1=1}^\infty
\frac{(-1)^{m_1}}{m_1}A_{m_1}, \slabel{b}\\
0&=&B_{m_2}\left[\ln\left(\frac{2}{\pi\epsilon}\right)+\frac{1}{m_2}+
\sum_{m_1=1}^\infty\left(\frac{2}{|{\bf m}|}C_{{\bf m}}-\frac{1}{m_1}
\right)\right] \nonumber\\
&&+\frac{(-1)^{m_2}}{m_2}A_0+2\sum_{m_1=1}^\infty\frac{(-1)^{m_1+m_2}}
{|{\bf m}|}C_{{\bf m}}A_{m_1}, \quad (m_2\geq 1). 
\end{subeqnarray}

\subsection{Determination of the slip length}

The flow at infinity is determined by substitution of Eq.~(\ref{veeAinf}) into
Eq.~(\ref{vee}), which gives
\begin{equation}\label{flowinf}
{\bf v}\sim\frac{2\pi U|z|}{b}{\bf e}_x\frac{1}{4}\int_{-1}^1
\int_{-1}^1[f_x(w,s)+f_y(w,s)]\d w\d s\quad  \hbox{as } |z|\to\infty,
\end{equation}
{whose simple form is due to the exponential decay of all Fourier modes except ${\bf m}={\bf 0}$.} 
Then Eqs.~(\ref{Fcoeffts}) and (\ref{ABeqn2n}) show that the shear rate at infinity is given by
\begin{equation}
\dot\gamma = \frac{\pi U (A_0+B_0)}{b},
\end{equation}
and therefore the slip length, $\lambda $, is found to be
\begin{equation}\label{slip}
\lambda \equiv \frac{U}{\dot\gamma}=\frac{b}{\pi(A_0+B_0)},
\end{equation}
which is proportional to the only other length scale in the problem, namely $b$.  {Notably, only the zeroth-order coefficients of Eqs.~(\ref{ABeqn3n}) contribute to the slip length}. The system given by Eqs.~(\ref{ABeqn3n}), independent 
of $b$, is solved in truncated form for various values of $\epsilon$ in \S\ref{numerics}.

\subsection{Error estimate}

An error estimate can be gleaned by retaining all terms in proceeding
from Eqs.~(\ref{veegrid}) to Eqs.~(\ref{inteqs}).  Although the former suggests 
that the symmetry of each grid element implies a mathematically even
dependence on $\epsilon$, convergence considerations prevent the error
bound from being $O(\epsilon^2)$.  For example, the Fourier cosine
series for $(\pi-x)^2$ in $[0,2\pi]$ shows that
\begin{equation}
\sum_{m=1}^\infty\frac{\sin(m\pi\epsilon S)\sin(m\pi\epsilon s)}
{\pi^2m^2}=\frac{\epsilon}{2}\min(s,S)[1-\epsilon\max(s,S)]
=O(\epsilon). 
\end{equation}
Hence the (relative) error estimate is $O(\epsilon\ln(1/\epsilon))$, as
in various sample calculations given by Leppington and Levine \cite{leppington72}. Mathematically, its
presence is due to an elliptic integral generated by the fundamental
singularity.

\section{Numerical calculation and analytical estimate of the mesh  slip length}
\label{numerics}

\begin{figure}[t]
\begin{center}
\includegraphics[width=0.8\textwidth]{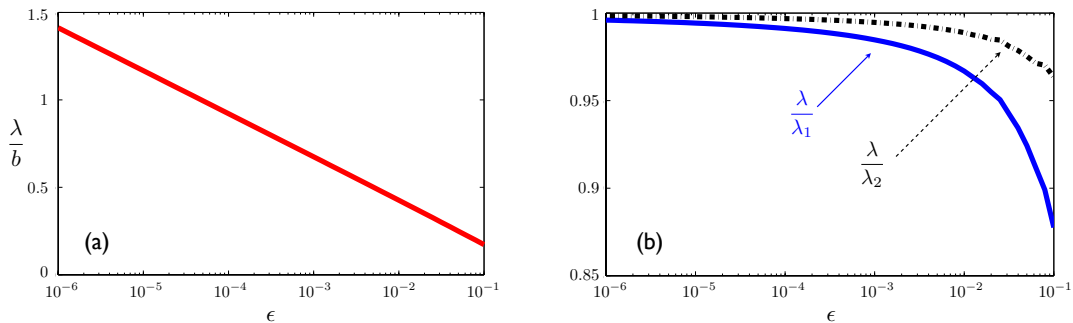}
\end{center}
\caption{(color online)  (a): Dimensionless effective slip length for the mesh-like super-hydrophobic surface, $\lambda / b$,  as a function of the  aspect ratio of the mesh, $\epsilon$. (b): Ratio between the slip length given by solving Eqs.~\eqref{ABeqn3n} and the simple estimates provided by Eq.~\eqref{Ldef} ($\lambda/\lambda_1$, blue, solid line) and Eq.~\eqref{2nd_order} ($\lambda/\lambda_2$, black, dash-dotted line), as a function of the aspect ratio.}
\label{results}
\end{figure}

\subsection{Numerical results}
\label{fit}
We now solve numerically  the infinite series given by Eqs.~\eqref{ABeqn3n} by truncating it at a finite value $m_1=m_2=N$. 
The numerical results are displayed in  Fig.~\ref{results}a, where we plot the dimensionless slip length, $\lambda/b$, as a function of the aspect ratio of the mesh, $\epsilon$, for a truncation size of $N=10^2$. The exact size of the truncation has little influence on the final computed results; for example, when  $\epsilon=10^{-3}$, the relative change in the computed slip length is less than 0.01 \% between $N=10^2$ and $N=10^3$.

Our results show that the surface does show a reduction in friction ({\it i.e.} $\lambda > 0$), and the effective slip length, always on the order of ({but typically smaller than}) the mesh size,  increases when the solid {area} fraction decreases. Since the data on Fig.~\ref{results}a are plotted on a semilog scale, we see that we recover the expected  logarithmic relationship between the effective  slip length of the mesh and  its aspect ratio  \cite{philip72,philip72_2,lauga03,ybert07}. A least squares fit to the data shown on Fig.~\ref{results}a  leads to the empirical formula
\begin{equation}\label{fit1}
{\frac{\lambda}{b}=\Lambda  \ln { \epsilon} + \Pi,}
\end{equation}
where the values  {$\Lambda=-0.107$ and  $\Pi=  -0.069$, gives result with maximum relative error of 0.4\%. Alternatively, if we define the solid area fraction of the surface $\phi_s$, it is easy to see that $\phi_s=\epsilon(2-\epsilon)$. A least-square fit of the type 
\begin{equation}
 \frac{\lambda}{b}=\Lambda_2  \ln {\phi_s} + \Pi_2,
\end{equation}
with the parameters $\Lambda_2=  -0.107$ and $\Pi_2= 0.003$, leads to a maximum relative error of $0.6\%$. As a matter of comparison, the fit proposed in Ref.~\cite{ng09_2}, based on numerical simulations for $\phi_s \gtrsim 0.05$, leads to a relative error of 5\% with the small-$\phi_s$ results obtained here; the asymptotic calculations presented in this paper are therefore able to quantitatively agree with the results of Ref.~\cite{ng09_2} which are valid for much larger mesh sizes.}

\subsection{Comparison with simple estimates}

An estimate of the mesh  slip length can be obtained analytically by performing the truncation by hand.  
If only $A_0,B_0$ --- that is the mean values of the density functions
with respect to $w$ --- are retained in Eqs.~(\ref{ABeqn3n}), the surviving two
equations give the lowest-order estimate
\begin{equation}\label{Ldef}
\frac{\lambda_1}{b}=\frac{1}{\pi (A_0+B_0)}=\frac{1}{3\pi }\ln\left(\frac{2}{\pi\epsilon}\right)\cdot
\end{equation}
We define here $L=\ln (2/\pi \epsilon)$ for subsequent convenience. 
{Note that the slope of the relationship given by Eq.~\eqref{Ldef} is $1/3\pi\approx 0.106$, which is very close to the fitted value $|\Lambda|=0.107$ in Eq.~\eqref{fit1}.}

Further, on writing Eqs.~(\ref{a}) and (\ref{b}) as
\begin{subeqnarray}
3&=&(A_0+B_0)\ln\left(\frac{2}{\pi\epsilon}\right)+\sum_{m=1}^\infty
\frac{(-1)^{m}}{m}(A_m+B_m), \\
-1&=&(A_0-B_0)\ln\left(\frac{2}{\pi\epsilon}\right)-\sum_{m=1}^\infty
\frac{(-1)^{m}}{m}(A_m-B_m), 
\end{subeqnarray}
and observing that the matrix elements in Eqs.~(\ref{ABeqn3n}) have $L$ only 
on the diagonal, it may be deduced that
\begin{equation}\label{slipcoefft}
\frac{\lambda_1}{b}-\frac{L}{3\pi}=O(L^{-1}).
\end{equation}
As an illustration of this result, the truncation of Eqs.~(\ref{ABeqn3n}) at
$m_1=m_2=1$ shows that the leading-order correction to Eq.~(\ref{Ldef}) leads to the improved estimate
\begin{equation}\label{2nd_order}
\frac{\lambda_2}{b}=\frac{L}{3\pi}-\frac{L+\frac{1}{2}}{6\pi\left[L^2+\frac{1}
{4}L-\frac{1}{2}+\frac{9}{4\sqrt{2}}(L+\frac{1}{2})\right]}\cdot
\end{equation}
On Fig.~\ref{results}b, we show the ratio between the actual slip length (as calculated in Fig.~\ref{results}a), and the simple estimates from Eq.~\eqref{Ldef} (blue, solid line) and Eq.~\eqref{2nd_order} (black, dash-dotted line). We see that these simple formulae  over-estimate the slip length (by up to 10\%), but asymptotically converge to the correct result when $\epsilon \to 0$. As expected,  the second-order solution, Eq.~\eqref{2nd_order}, is quantitatively better than the first-order one, Eq.~\eqref{Ldef}.

\section{Conclusion}

In this work we have considered super-hydrophobic surface in the shape of a square mesh, and have presented an analytical method to calculate their effective slip length. {Our analysis, which is valid for asymptotically small aspect ratio of the mesh, agrees also  quantitatively with numerical calculations valid for larger mesh width~\cite{ng09_2}.
The results we obtain show that, for thin meshes, such surfaces can provide slip lengths on the order of the mesh size.  However, for realistic mesh aspect ratio,  say  10\% or larger, the slip length is only a fraction of the typical mesh length (see Fig.~\ref{results}).} The increase of the slip length with the solid-to-air ratio of the overall surface is logarithmic, a universal feature of super-hydrophobic coatings with long and thin no-slip domains  \cite{philip72,philip72_2,lauga03,ybert07}. Beyond such physical scaling, our  results {provide a first  analytical prediction} for the resistance to shear flow of the recently devised super-hydrophobic surfaces made of metal wires \cite{feng04,sun05}, and more generally to geometries with grid-like features such as fabric and textiles composed of interwoven threads.

Although we have considered here an idealized model system, the analytical approach  developed in the paper has the advantage that it provides the effective slip length of the surface with relative error of order $\epsilon \log (1/\epsilon)$, which   is significantly better than the more traditional but only   logarithmically correct
slender-body approach.

\section*{Acknowledgments}
We thank Shutao Wang for allowing us to reproduce in Fig.~\ref{grid} the picture from Ref.~\cite{wang07}.
This work was supported in part by the National Science Foundation (Grants No. CTS-0624830 and CBET-0746285).


\end{document}